\documentstyle[12pt,preprint]{aastex}

\newcommand{\xray}{\mbox{X-ray }}
\newcommand{\hst}{\textit{HST }}
\newcommand{\fgs}{\textit{FGS }}

\begin{document}

\title{\Large\bf A single sub-km Kuiper Belt object from a stellar
Occultation in archival data} 

\author{H.~E.~Schlichting$^{1,2}$, E.~O.~Ofek$^{1,3}$, M.~Wenz$^4$,
  R.~Sari$^{1,5}$, \\
A.~Gal-Yam$^6$, M.~Livio$^7$, E.~Nelan$^7$, S.~Zucker$^8$}

\maketitle

\noindent{\scriptsize $^1$ Department of Astronomy, 249-17, California Institute of Technology,
Pasadena, CA 91125, USA\\
$^2$ CITA, University of Toronto, 60 St. George St., ON, M5S 3H8, Canada\\
$^3$ Einstein Fellow\\
$^4$ Goddard Space Flight Center, 8800 Greenbelt Road, Greenbelt, MD 20771,
USA\\
$^5$ Racah Institute of Physics, Hebrew University, Jerusalem 91904, Israel\\
$^6$ Faculty of Physics, Weizmann Institute of Science, POB 26, Rehovot 76100,
Israel\\
$^7$ Space Telescope Science Institute, 3700 San Martin Drive, Baltimore, MD
21218, USA\\
$^8$ Department of Geophysics and Planetary Sciences, Tel Aviv University, Tel Aviv 69978, Israel\\}

{\small \bf The Kuiper belt is a remnant of the primordial Solar
  System. Measurements of its size distribution constrain its accretion and
  collisional history, and the importance of material strength of Kuiper belt
  objects (KBOs)\cite{DF97,SC97,KL99,PS05}. Small, sub-km sized, KBOs elude
  direct detection, but the signature of their occultations of background
  stars should be detectable\cite{B76,D92,AAC92,RMS87,ZBL08}. Observations at
  both optical\cite{RDD06} and X-ray\cite{CKL06} wavelengths claim to have
  detected such occultations, but their implied KBO abundances are
  inconsistent with each other and far exceed theoretical expectations. Here,
  we report an analysis of archival data that reveals an occultation by a body
  with a $\sim$500\,m radius at a distance of 45\,AU. The probability of this
  event to occur due to random statistical fluctuations within our data set is
  about 2\%. Our survey yields a surface density of KBOs with radii larger
  than 250\,m of $2.1^{+4.8}_{-1.7}\times10^7\,\rm{\deg^{-2}}$, ruling out
  inferred surface densities from previous claimed detections by more than
  5\,$\sigma$. The fact that we detected only one event, firmly shows a
  deficit of sub-km sized KBOs compared to a population extrapolated from
  objects with $r>50$\,km. This implies that sub-km-sized KBOs are undergoing
  collisional erosion, just like debris disks observed around other stars.}\\

A small KBO crossing the line of sight to a star will partially obscure the
stellar light, an event which can be detected in the star's light curve. For
visible light, the characteristic scale of diffraction effects, known as the
Fresnel scale, is given by $(\lambda a/2)^{1/2} \sim 1.3\,\rm{km}$, where
$a\sim 40\,\rm{AU}$ is the distance to the Kuiper belt and $\lambda \sim
600\,\rm{nm}$ is the wavelength of our observations. 

Diffraction effects will be apparent in the star's light curve due to
occulting KBOs provided that both star and the occulting object are smaller
than the Fresnel scale \cite{RM00,NL07}. Occultations by objects smaller than
the Fresnel scale are in the Fraunhofer regime. In this regime the diffraction
pattern is determined by the size of the KBO and its distance to the observer,
the angular size of the star, the wavelength range of the observations and the
impact parameter between the star and the KBO (see Supplementary Information
for details). The duration of the occultation is approximately given by the
ratio of the Fresnel scale to the relative velocity perpendicular to the line
of sight between the observer and the KBO. Since the relative velocity is
usually dominated by the Earth's velocity around the Sun, which is
$30\,\rm{km\,s^{-1}}$, typical occultations only last of order of a tenth of a
second.

Extensive ground based efforts have been conducted to look for optical
occultations \cite{RDD06,ZBL08,BKW08,BP09}. To date, these visible searches
have announced no detections in the region of the Kuiper belt (30-60\,AU), but
one of these quests claims to have detected some events beyond 100\,AU and at
about 15\,AU \cite{RDD06}. Unfortunately, ground based surveys may suffer from
a high rate of false-positives due to atmospheric scintillation, and lack the
stability of space based platforms. The ground breaking idea to search for
occultations in archival RXTE \xray data resulted in several claimed
occultation events \cite{CKL06}. Later, revised analysis of the \xray data
\cite{Chang07,JLM08,LC08,BPA09} conclude that the majority of the originally
reported events are most likely due to instrumental dead time effects. Thus,
previous reports of optical and \xray events remain dubious \cite{BKW08} and
their inferred KBO abundance is inconsistent with the observed break in the
KBO size distribution, which has been obtained from direct detections of large
KBOs \cite{BTA04,FH108,FK108}. Furthermore, they are also difficult to
reconcile with theoretical expectations, which predict collisional evolution
for KBOs smaller than a few km in size \cite{D69,PS05} and hence a lower KBO
abundance than inferred from extrapolation from KBOs with $r > 50$\,km.

For the past 14 years, the Fine Guidance Sensors (\textit{FGS}) on board of
{\it Hubble Space Telescope} (\textit{HST}) have been collecting photometric
measurements of stars with 40\,Hz time resolution, allowing for the detection
of the occultation diffraction pattern rather than a simple decrease in the
photon count. We examined four and a half years of archival \fgs data, which
contain $\sim 12,000$ star hours of low ecliptic latitude ($\vert b \vert <
20^{\circ}$) observations.

Our survey is most likely to detect occultations by KBOs that are 200-500\,m
in radius given the signal-to-noise of our data (Supplementary Figure 3) and a
power-law size distribution with power-law index between 3 and
4.5. Occultation events in this size range are in the Fraunhofer regime where
the depth of the diffraction pattern varies linearly with the area of the
occulting object and is independent of its shape. The theoretical light curves
for our search algorithm were therefore calculated in this regime. We fitted
these theoretical occultation templates to the FGS data and performed $\chi^2$
analysis to identify occultation candidates (see Supplementary Information).
We detected one occultation candidate, at ecliptic latitude $14^{\circ}$, that
significantly exceeds our detection criterion (Figure~1). The best fit
parameters yield a KBO size of $r=520 \pm 60\,\rm{m}$ and a distance of
$45^{+5}_{-4}$\,AU where we assumed a circular KBO orbit and an inclination of
$14^{\circ}$. Using bootstrap simulations, we estimate a probability of $\sim
2\%$ that such an event is caused by statistical fluctuations over the whole
analyzed \fgs data set (Supplementary Figure 7). We note that for objects on
circular orbits around the sun two solutions can fit the duration of the
event. However, the other solution is at a distance of 0.07~AU from the Earth,
and is therefore unlikely. It is also unlikely that the occulting object was
located in the Asteroid belt, since the expected occultation rate from
Asteroids is about two orders of magnitude less than our implied
rate. Furthermore, an Asteroid would have to have an eccentricity of order
unity to be able to explain the duration of the observed occultation event.

Using the KBO ecliptic latitude distribution from Elliot et. al (2005)
\cite{E05}, our detection efficency, and our single detection, we constrain
the surface density around the ecliptic (averaged over $-5^{\circ} < b <
5^{\circ}$) of KBOs with radii larger than 250\,m to $2.1^{+4.8}_{-1.7} \times
10^7\,\rm{\deg^{-2}}$ (see Supplementary Information Sections 5 and 6). This
surface density is about three times the implied surface density at
$5.5^{\circ}$ ecliptic latitude and about five times the surface density at
$8-20^{\circ}$ ecliptic latitude. This is the first measurement of the surface
density of hecto-meter-sized KBOs and it improves previous upper limits by
more than an order of magnitude \cite{ZBL08,BP09}. Figure~2 displays our
measurement for the sub-km KBO surface density and summarizes published upper
limits from various surveys. Our original data analysis focused on the
detection of KBOs located at the distance of the Kuiper belt between 30\,AU
and 60\,AU. In order to compare our results with previously reported
ground-based detections beyond 100\,AU \cite{RDD06}, we performed a second
search of the \fgs data that was sensitive to objects located beyond the
classical Kuiper belt. Our results challenge the reported ground-based
detections of two 300\,m-sized objects beyond 100\,AU \cite{RDD06}. Given our
total number of star hours and a detection efficency of 3\% for 300\,m-sized
objects at $\sim 100\,\rm{AU}$ we should have detected more than twenty
occultations. We therefore rule out the previously claimed optical detections
\cite{RDD06} by more than 5\,$\sigma$. This result accounts for the broad
latitude distribution of our observations (i.e., $\vert b \vert < 20^{\circ}$)
and the quoted detection efficency of our survey includes the effect of the
finite angular radii of the guide stars at 100\,AU.

The KBO cumulative size distribution is parameterized by $N(>r)\propto
r^{1-q}$, where $N(>r)$ is the number of objects with radii greater than $r$,
and $q$ is the power-law index. The power-law index for KBOs with radii above
$\sim$45\,km is $\sim 4.5$ \cite{FH108,FK108} and there is evidence for a
break in the size distribution at about $r_{\rm{break}} \sim 45\,\rm{km}$
\cite{BTA04,FH108,FK108}. We hence use this break radius and assume a surface
density for KBOs larger than $r_{\rm{break}}$ \cite{FH08} of $5.4\, \deg^{-2}$
around the ecliptic. Accounting for our detection efficency, the velocity
distribution of the \hst observations, and assuming a single power-law for
objects with radii less than 45\,km in size, we find
$q=3.9^{+0.3,+0.4}_{-0.3,-0.7}$ (1 and 2\,$\sigma$ errors) below the
break. Our results firmly show a deficit of sub-km sized KBOs compared to
large objects. This confirms the existence of the previously reported break
and establishes a shallower size distribution extending two orders of
magnitude in size down to sub-km sized objects. This suggests that sub-km
sized KBOs underwent collisional evolution, eroding the smaller KBOs. This
collisional grinding in the Kuiper belt provides the missing link between
large KBOs and dust producing debris disks around other stars. Currently our
results are consistent with a power-law index of strength dominated
collisional cascade \cite{D69}, $q=3.5$, within $1.3\sigma$ and with
predictions for strengthless rubble piles \cite{PS05}, $q=3.0$, within
$2.4\sigma$. An intermediate value of $3<q<3.5$ implies that KBOs are
strengthless rubble piles above some critical size, $r_c < r < 45\,\rm{km}$,
and strength dominated below it, $r < r_c $. Our observations constrain for
the first time $r_c$. At the 2\,$\sigma$ level we find $r_c > 3\,\rm{km}$.

Using our estimate for the size distribution power-law index ($q=3.9$) and our
KBO surface density for 250\,m sized KBOs at an ecliptic latitude of
$b=5.5^{\circ}$, which is the ecliptic latitude of the RXTE observations of
Scorpius X-1, we predict that there should be $\sim 3.6 \times 10^{9}$
30\,m-radius objects per square degree. This is about 150 times less than the
original estimate from \xray observations of Scorpius~X-1 that reported 58
events \cite{CKL06}, and it is about 30 times less than the revised estimate
from the same \xray observations, which concludes that up to 12 events might
be actual KBO occultations \cite{Chang07}. Our results rule out the implied
surface density from these 12 events at 7\,$\sigma$ confidence level. One can
reconcile our results and the claimed \xray detections only by invoking a
power-law index of $q \sim 5.5 $ between 250\,m and 30\,m. More recent \xray
work reports no new detections in the region of the Kuiper belt but places an
upper limit of $1.7 \times 10^{11}\,\rm{\deg^{-2}}$ for objects of 50\,m in
radius and larger \cite{LC08}. This is consistent with the KBO surface density
of $N(>50\,m)=8.2 \times 10^8\,\rm{\deg^{-2}}$ that we derive by extrapolating
from our detection in the hecto-meter size range.

The statistical confidence level on our detection is 98\%. However, our
conclusions that there is a significant break in the size distribution and
that collisional erosion is taking place and the significant discrepancy with
previously claimed occultation detections rely on the {\it low number} of
events we discovered. These conclusions would only be strengthened if this
event was caused by an unlikely statistical fluctuation or a yet unknown
instrumental artifact.

Ongoing analysis of the remaining \fgs data, which will triple the number of
star hours, together with further development of our detection algorithm
(i.e., including a larger number of light curve templates) holds the promise
for additional detections of occultation events and will allow us to
constrain the power-law index of the size distribution further.


\noindent{\bf Acknowledgments } We thank Dr. H. K. Chang for valuable comments
that helped to improve this manuscript. Some of the numerical calculations
presented here were performed on Caltech's Division of Geological and
Planetary Sciences Dell cluster. Partial support for this research was
provided by NASA through a grant from the Space Telescope Science
Institute. R. S. acknowledges support from the ERC and the Packard
Foundation. A. G. is supported by the Israeli Science Foundation, an EU
Seventh Framework Programme Marie Curie IRG fellowship and the Benoziyo Center
for Astrophysics, a research grant from the Peter and Patricia Gruber Awards,
and the William Z. and Eda Bess Novick New Scientists Fund at the Weizmann
Institute. S. Z. acknowledges support from the Israel Science Foundation --
Adler Foundation for Space Research.\\

\noindent{\bf Author Contributions } H. E. S. wrote the detection algorithm,
analyzed the \fgs data for occultation events, calculated the detection
efficency of the survey, preformed the bootstrap analysis and wrote the
paper. E. O. O. calculated the stellar angular radii, the velocity information
of the observations, the correlated noise and other statistical properties of
the data. R. S. guided this work and helped with the scientific interpretation
of the results. A.G. proposed using \hst\fgs data for occultation studies and
helped to make the data available for analysis. M. W. extracted the \fgs
photometry streams and provided coordinates and magnitudes of the guide
stars. M. L. helped in gaining access to the \fgs data and provided insights
into the operation and noise properties of the \fgs. E. N. provided expert
interpretation of the FGS photometric characteristics in the HST operational
environment. S. Z. took part in the statistical analysis of the data. All
authors discussed the results and commented on the manuscript.\\

\noindent{\bf Author Information } Reprints and permissions information is
  available at www.nature.com/reprints. Correspondence and requests for
  materials should be addressed to H. E. S. (hes@astro.caltech.edu) or
  E. O. O. (eran@astro.caltech.edu).\\

\begin{figure} [htp]
\centerline{\includegraphics[ scale=1.4]{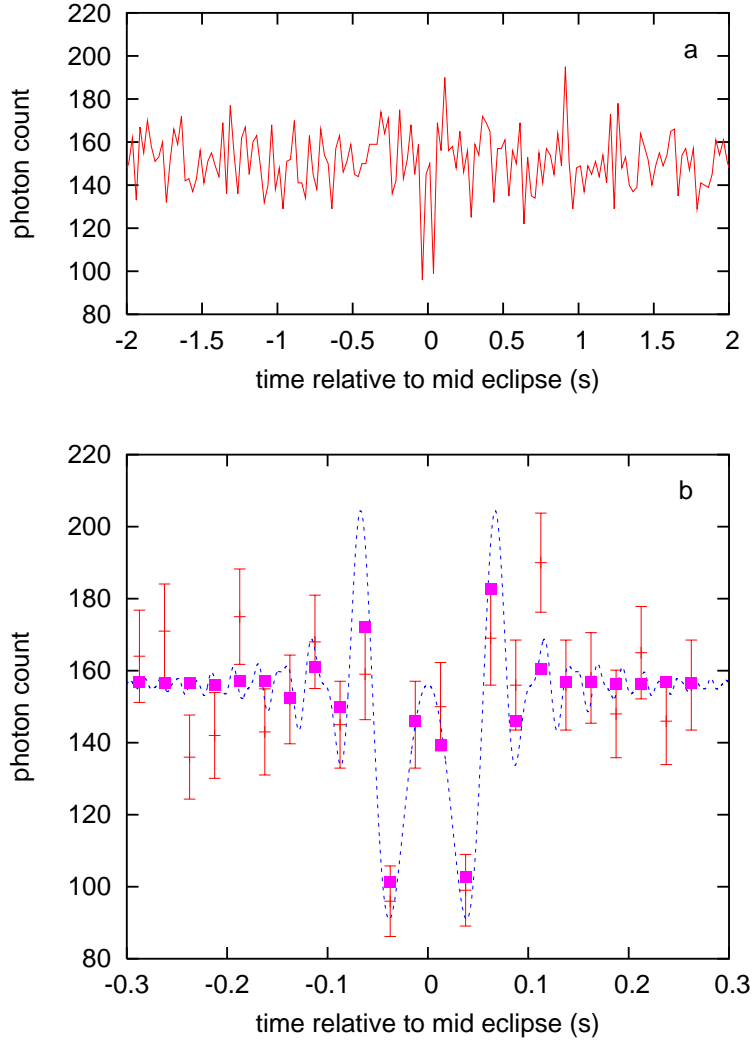}}
\caption{\footnotesize Photon counts as a function of time of the
candidate occultation event observed by FGS2. Part a) shows the photon count
spanning $\pm 2$\,seconds around the occultation event. Part b) displays the
event in detail. The red crosses and error bars are the \fgs data points with
Poisson error bars, the dashed, blue line is the theoretical diffraction
pattern (calculated for the 400-700\,nm wavelength range of the \fgs
observations), and the pink squares correspond to the theoretical light curve
integrated over 40\,Hz intervals. Note, the actual noise for this observation
is about 4\% larger than Poisson noise due to additional noise sources such as
dark counts (about 3 to 6 counts in a 40\,Hz interval), and jitter due to the
displacement of the guide star (by up to 10\,mas) from its null position. The
mean signal-to-noise ratio in a 40\,Hz interval for the roughly half an hour
of observations is $\sim 12$. The event occurred at UTC 05:17:49 2007, Mar
24. The best fit $\chi^2/\rm{dof}$ is $20.1/21$. The star has an ecliptic
latitude of +14. Its angular radius and effective temperature are $\approx
0.3$ of the Fresnel scale and $\approx 4460\,\rm{K}$, respectively. These
values were derived by fitting the 2MASS \cite{SCS06} JHK and USNO-B1 BR
\cite{M03} photometry with a black-body spectrum. The position of the star is
R.A.=186.87872$^\circ$, Dec=12.72469$^\circ$ (J2000) and its estimated
V-magnitude is 13.4. The auto-correlation function (excluding lag zero) of the
photometric time series of this event is consistent with zero within the
statistical uncertainty. Each \fgs provides two independent PMT readings and
we confirmed that the occultation signature is present in both of these
independent photon counts. We examined the photon counts of the other guide
star that was observed by FGS1 at the time of the occultation and confirmed
that the occultation signal is only present in the observations recorded by
FGS2. We examined the engineering telemetry for \hst around the time of the
event and verified that the guiding performance of \hst was normal. We
therefore conclude that the above occultation pattern is not caused by any
known instrumental artifacts.}
\end{figure}

\begin{figure} [htp]
\centerline{\includegraphics[ scale=1.6]{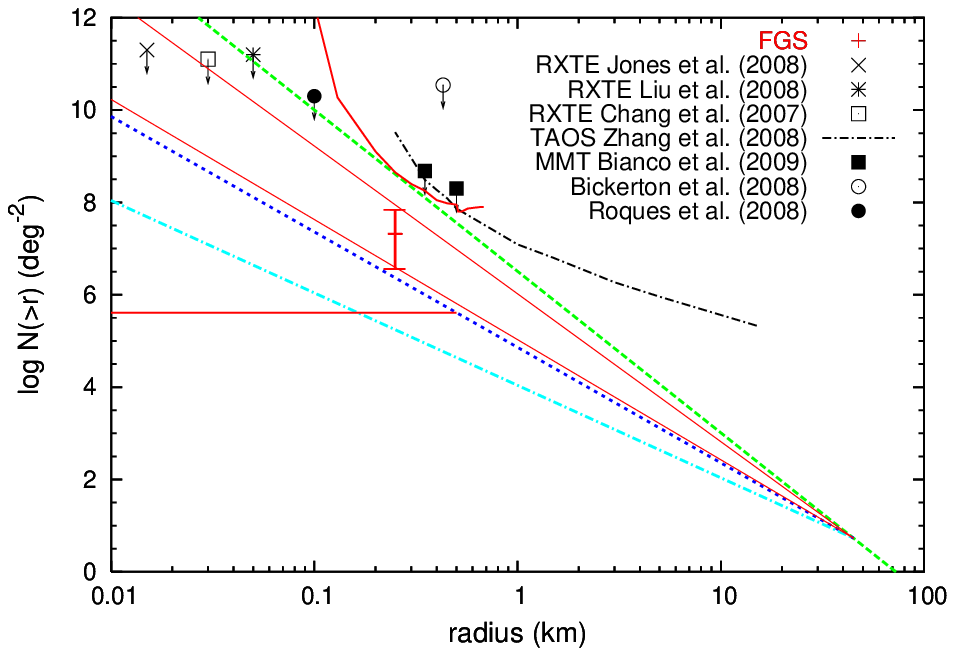}}
\caption{\footnotesize Cumulative KBO size distribution as a function
 of KBO radius for objects located between 30 and 60\,AU. The results from our
 \fgs survey are shown in red and are presented in three different ways: (i)
 The cross is derived from our detection and represents the KBO surface
 density around the ecliptic (averaged over $ -5^{\circ} <b < 5^{\circ}$) and
 is shown with 1\,$\sigma$ error bars. The cross is plotted at
 $r=250\,\rm{m}$, which is roughly the peak of our detection probability (see
 Supplementary Information Section 6 for details). (ii) The upper and lower
 red curves correspond to our upper and lower 95\% confidence level which were
 derived without assuming any size distribution. (iii) The region bounded by
 the two straight red lines falls within $1\sigma$ of our best estimate for
 the power-law size distribution index, i.e. $q=3.9 \pm 0.3$, which was
 calculated for low ecliptic latitudes ($\vert b \vert <5^{\circ}$). These
 lines are anchored to the observed surface density at $r=45\,\rm{km}$. For
 comparison, the green (long-dashed) line is the observed size distribution of
 large KBOs (i.e., $r>45\,\rm{km}$), which has $q=4.5$, extrapolated as a
 single power-law to small sizes. The blue (short-dashed) line is a double
 power-law with $q=3.5$ (collisional cascade of strength dominated bodies) for
 KBOs with radii less than 45\,km and $q=4.5$ above. The cyan (dot-dashed)
 line corresponds to $q=3.0$ (collisional cascade of strengthless rubble
 piles) for KBOs below 45\,km in size. All distributions are normalized to
 $N(>r)=5.4\,\rm{deg^{-2}}$ at a radius of 45\,km \cite{FH08}. In addition,
 95\% upper limits from various surveys are shown in black. Note, a power-law
 index of 3.9 was used for calculating the cumulative KBO number density from
 the RXTE observations.}
\end{figure}

\clearpage

\underline{\Large Supplementary Information}\\
\\
{\bf 1 The FGS data set}
\\
\\
There are three \fgs on board of \hst. Each \fgs consists of four
photomultipliers (PMTs). Nominal \hst operation uses two \fgs for guiding,
with each \fgs observing its own guide star. The photon counts recorded by
each \fgs are therefore different, but global instrumental artifacts and
Observatory level transients will display in both \fgs and can therefore be
identified and removed.

Observations of the inclination distribution of large KBOs find that about
75\% have an inclination angle $\vert i \vert \lesssim 20^{\circ}$
\cite{J96,BT01,E052}. We therefore divide the \fgs observations into a low
ecliptic latitude ($\vert b \vert < 20^{\circ}$) and a high ecliptic latitude
($\vert b \vert > 20^{\circ}$) sample. The high-ecliptic latitude observations
($\vert b \vert>20^{\circ}$) provide an excellent control sample.\\
\\

{\bf 2 \fgs Guide Stars} 
\\
\\
The \fgs guide stars span a broad range of magnitudes and spectral
types. The signal-to-noise ratio, S/N, in a 1/40\,s data bin depends on the
magnitude of the star. Its distribution is shown in Supplementary Figure
\ref{SI_fig1}.

The angular sizes of guide stars were derived by fitting the 2MASS
\cite{SCS062} JHK and USNO-B1 BR \cite{M032} photometry with a black-body
spectrum. Supplementary Figure \ref{SI_fig2} shows the angular radii
distribution of the guide stars. About 66\% of the stars in our data set
subtend angular sizes less than 0.5 of the Fresnel scale at a distance of
40\,AU. The diffraction pattern that is produced by a sub-km sized KBO
occulting an extended background star is smoothed over the finite stellar
disk. This effect becomes clearly noticeable for stars that subtend sizes
larger than about 0.5 of a Fresnel scales \cite{RM002,NL072} and it reduces the
detectability of occultation events around such stars. The effect of finite
angular radii of the guide stars on the detection efficiency of our survey is
taken into account (see Detection Efficency section 5 for details).\\
\\

{\bf 3 Detection Algorithm}
\\
\\
Our detection algorithm performs a template search with theoretical
light curves and uses a $\chi^2$ fitting procedure to identify occultation
candidates. Our survey is most likely to detect KBO occultation events caused
by objects that are 200-500\,m in radius given the signal-to-noise of our data
and for a power-law index of the KBO size distribution, $q$, between 3 and
4.5. Occultation events in this size range are in the Fraunhofer regime. The
theoretical light curves for our search algorithm are therefore calculated in
the Fraunhofer regime. Our templates are calculated for various impact
parameters assuming a point source background star and are integrated over the
400-700\,nm wavelength range of the \fgs observations. For a given impact
parameter between the KBO and the star, our theoretical light curves have
three free parameters that we fit for. The first is the mean number of photon
counts, which is the normalization of the light curve. The second is the
amplitude of the occultation, which is proportional to the size of the KBO,
and the third is the width of the occultation, which is independent of the
object size, and is determined by the ratio of the Fresnel scale to the
relative speed between \hst and the KBO perpendicular to the line of
sight. This relative speed is determined by the combination of \hst's
velocity around the Earth, Earth's velocity around the Sun and the velocity of
the KBO itself. We use this information to restrict the parameter space for
the template widths in our search such that we are sensitive to KBOs located
at the distance of the Kuiper belt between 30\,AU and 60\,AU.\\
\\

{\bf 4 Detection Criterion and Significance Estimates} 
\\
\\
The significance of occultation candidates can be measured by their $\Delta
\chi^2$ which is defined here as the difference between the $\chi^2$
calculated for the best fit of a flat light curve, which corresponds to no
event, and the $\chi^2$ of the best fit template. Occultation events have
large $\Delta \chi^2$, since they are poorly fit by a constant. Cosmic ray
events, which give rise to one very large photon count reading in a 40\,Hz
interval, can also result in a large $\Delta \chi^2$ but the fit of the
occultation template is very poor. We examined all flagged events for which
the template fit of the diffraction pattern was better than
15\,$\sigma$. About a handful of false-positives where flagged by our
detection algorithm that have a value of $\Delta \chi^2$ comparable to or
larger than the occultation event. However, in {\it all} cases these
false-positives were caused by a 1~Hz jitter due to the displacement of the
guide star from its null position. The occultation event itself did not show
any such jitter. To determine the $\Delta \chi^2$ detection criterion for our
search algorithm and to estimate the probability that detected events are due
to random noise we use the bootstrap technique \cite{E82}. Specifically, from
a given \fgs time series of length N we randomly selected N points with
repetitions and created `artificial' time series from it. We analyzed these
`artificial' data sets using the same search algorithm that we applied to the
actual \fgs data. This technique creates random time series with noise
properties identical to those of the actual data, but it will lose any
correlated noise. Therefore, this technique is justified if there is no
correlated noise in the data sets. To look for correlated noise we calculated
the auto-correlation function, with lags between 0 to 1\,s. Most of the data
sets are free of statistical significant correlated noise. The $\sim$ 12\% of
the data sets that did show correlated noise exceeding 4\,$\sigma$, which was
often due to slopes (e.g., long-term variability) in the data sets, were
excluded from the bootstrap analysis.

The \fgs data set consists of observations of many different stars with
magnitudes ranging from 9 to 14. The number of photon counts and
signal-to-noise properties vary therefore from observation to observation (see
Supplementary Figure \ref{SI_fig1} for the signal-to-noise ratio distribution
of the FGS observations). Our $\Delta \chi^2$ calculation accounts for the
Poisson noise of the data. Therefore, the probability that occultation
candidates are due to random noise can be characterized by a single value of
$\Delta \chi^2$ for all observations, irrespective of the mean photon count of
a given observation provided that the noise properties across all observations
are well characterized by a Poisson distribution. In reality, the noise
properties are different from observation to observation; especially
non-Poisson tails in the photon counts distribution will give rise to
slightly different $\Delta \chi^2$ distributions. Therefore, ideally, we would
determine a unique detection criterion for each individual data set. However,
this would require to simulate each data set, which contains about an hour of
observations in a single \hst orbit, over the entire length of our survey
($\sim 12,000$ star hours). This is not feasible due to the enormous
computational resources that would be required, i.e. simulating a single one
hour data set over the entire survey length requires about 5\,CPU days, which
corresponds to $\sim 60,000$\,CPU days for the entire \fgs survey. Instead,
we perform the bootstrap simulation over all the \fgs data sets together,
where each individual data set was simulated about a 100 times, which required
about $\sim 500$\,CPU days in total. This way we estimate the typical $\Delta
\chi^2$ value that corresponds to having less than one false-positive
detection over the $\sim 12,000$ star hours of low ecliptic observations. For
all occultation candidates that exceed this detection threshold, we determined
their statistical significance, i.e. the probability that they are due to
random noise, by extensive bootstrap simulations of the individual data sets
(Supplementary Figure \ref{SI_fig5}).\\
\\

{\bf 5 Detection Efficency}
\\
\\
The ability to detect an occultation event of a given size KBO depends on the
impact parameter of the KBO, the duration of the event, the angular size of
the star and the signal-to-noise ratio of the data. We determined the
detection efficiency of our survey by recovering synthetic events that we
planted into the observed photometric time series by multiplying the actual
\fgs data with theoretical light curves of KBO occultation events. The
synthetic events correspond to KBO sizes ranging from $130\,\rm{m} < r <
650\,\rm{m}$, they have impact parameters from 0 to 5.5 Fresnel scales and a
relative velocity distribution that is identical to that of the actual \fgs
observations. To account for the finite angular sizes of the stars we
generated light curve templates with stellar angular radii of 0.1, 0.2, 0.3,
0.4, 0.6, 0.8 and 1 Fresnel scales distributed as shown in Supplementary
Figure \ref{SI_fig2}. The modified light curves with the synthetic events were
analyzed using the same search algorithm that we used to analyze the \fgs
data. The detection efficency of our survey was calculated using the angular
size distribution of the \fgs guide stars assuming a distance of 40\,AU. We
normalize our detection efficency for a given size KBO, $\eta(r)$ , to 1 for
an effective detection cross section with a radius of one Fresnel scale.

The detection efficiency of our survey is $\sim 0.05$ ($\sim 0.6$) for objects
with $r=200\,\rm{m}$ ($r=500\,\rm{m}$) located at 40\,AU. Note that this value
for the detection efficency already accounts for the angular radii
distribution of the guide stars (e.g., for comparison, stars that subtend
angular radii less than 0.5 of the Fresnel scale result in a detection
efficency of $\sim 0.08$ [$\sim 0.8$] for objects with $r=200\,\rm{m}$
[$r=500\,\rm{m}$].).\\
\\

{\bf 6 Calculating the KBO Surface Density}
\\
\\
The number of occultation events is given by
\begin{equation}
N_{events}\simeq -2 v_{rel} F \int_{r_{min}}^{r_{max}} \int_{-b}^{b}\eta(r)
  \frac{\Delta t}{\Delta b} \frac{dN(r,b)}{dr} \,\mathrm{d}b
  \mathrm{d}r
\label{eq1}
\end{equation}
where $v_{rel}=23\,\rm{km/s}$ is the typical relative velocity between the KBO
and the observer, $b$ is the ecliptic latitude, $\Delta t/\Delta b$ is the
time observed per degree in ecliptic latitude (see Supplementary Figure
\ref{SI_fig3}) and $F=1.3\rm{\,km}$ is the Fresnel scale. The number density
of KBOs is both a function of ecliptic latitude and the KBO radius, $r$. Here
we assume that the KBO latitude distribution, $f(b)$, is independent of size
and we take the distribution provided in Elliot et al. (2005) \cite{E052}. We
further assume that the KBO size distribution follows a power law. It can
therefore be written as $N(r,b)=n_0 \times r^{-q+1} \times f(b)$ where $n_0$
is the normalization factor for the cumulative surface density of
KBOs. Substituting for $dN(r,b)/dr$ in equation \ref{eq1} and solving for
$n_0$ we have
\begin{equation}
n_0\simeq \frac{N_{events}}{2 v_{rel} F (q-1) \int_{r_{min}}^{r_{max}}
  \eta(r) r^{-q} \,\mathrm{d}r \int_{-b}^{b}
  f(b)\frac{\Delta t}{\Delta b} \,\mathrm{d}b}.
\label{eq3}
\end{equation}
Evaluating equation \ref{eq3} yields a cumulative KBO surface density averaged
over the ecliptic ($\vert b \vert < 5^{\circ}$) of
\begin{equation}
N(r>250\,m) \simeq 2.1 \times 10^7\,\rm{deg^{-2}}
\end{equation}
We assumed $q=4$ when evaluating the integral over $r$. We note however that
the value for the cumulative KBO surface density at $r=250\,\rm{m}$ only
depends weakly on the exact choice for $q$ [e.g. $N(r>250\,m)$ only
ranges from $2.3 \times 10^7\,\rm{deg^{-2}}$ to $2.1 \times
10^7\,\rm{deg^{-2}}$ for values of $q$ between 3 and 4.5]. We quote our
results as the KBO surface density of objects larger than 250\,m in radius
since this is roughly the size of KBOs, which our survey is most likely to
detect given our detection efficency and a power-law size distribution with
$q=3-4.5$. The implied surface density for KBOs with radii larger than 250\,m
is $7.7 \times 10^6\,\rm{deg^{-2}}$ at $b=5.5^{\circ}$, which is the ecliptic
latitude of the RXTE observations of Scorpius X-1, and it is $4.4 \times
10^6\,\rm{deg^{-2}}$ for $8^{\circ}<\vert b \vert < 20^{\circ}$.\\

\begin{figure} [htp]
\centerline{\includegraphics[ scale=1.7]{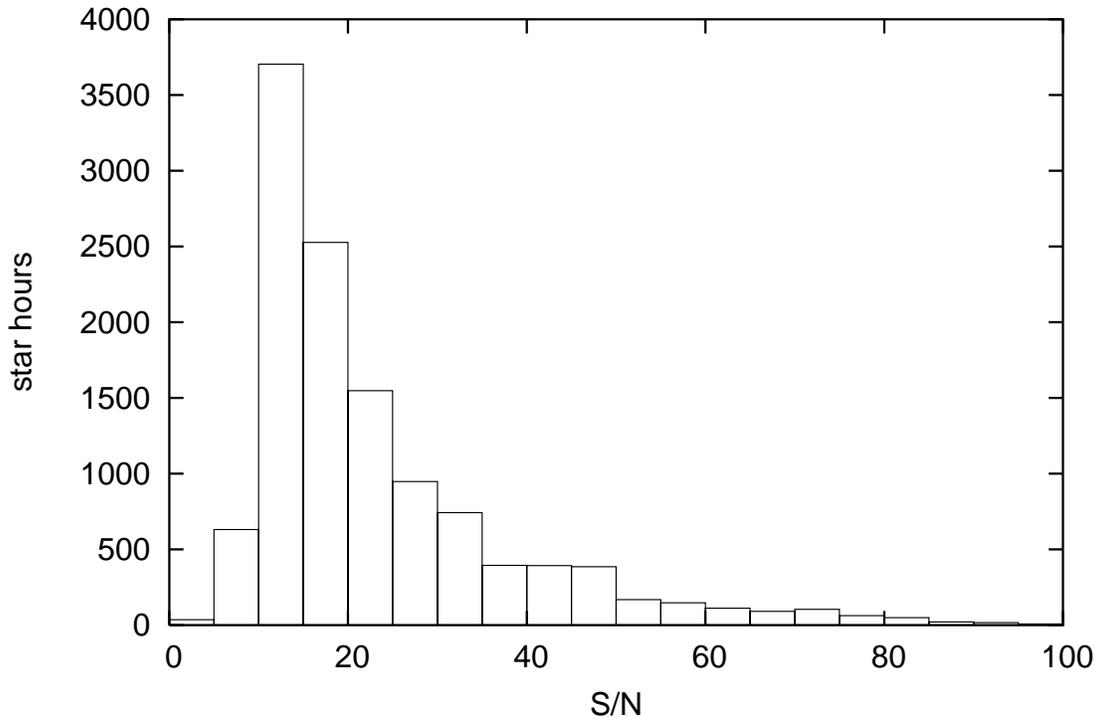}}
\caption{\footnotesize Distribution of star hours as a function of the mean
  signal-to-noise ratio, $S/N$, in a 40\,Hz bin for the 12,000 hours of low
  ecliptic latitude observations ($\vert b \vert < 20^{\circ}$) in the
  analyzed FGS data set.}
\label{SI_fig1}
\end{figure}

\begin{figure} [htp]
\centerline{\includegraphics[ scale=1.7]{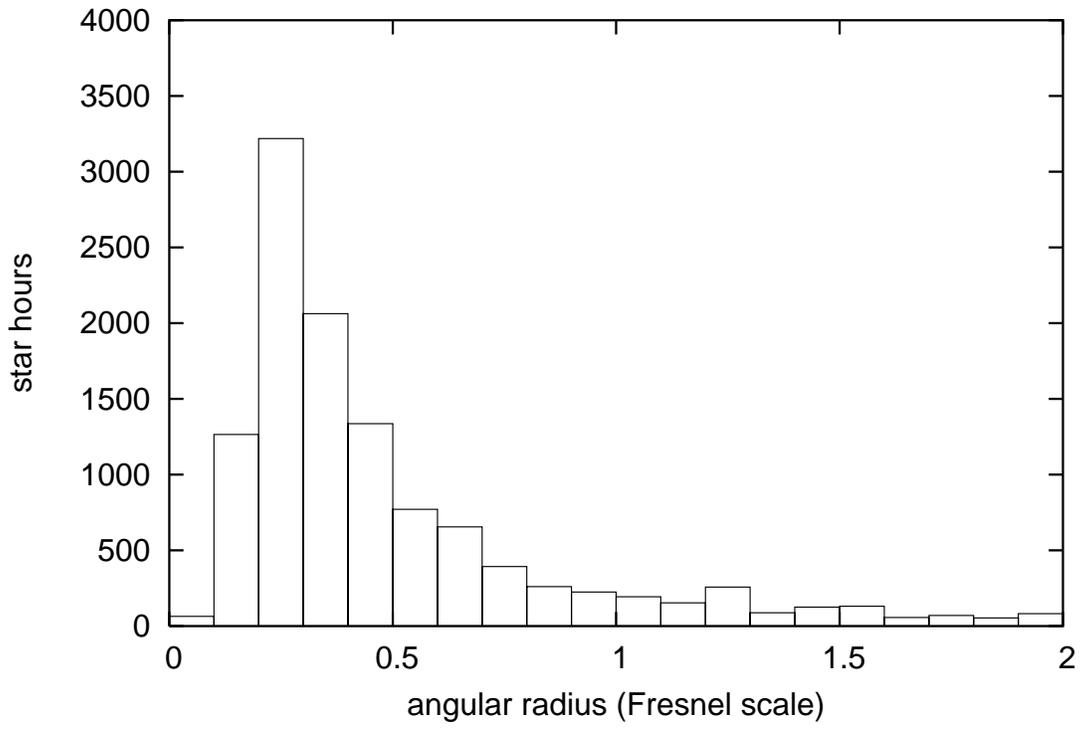}}
\caption{\footnotesize Distribution of star hours as a function of
angular radii of the guide stars. The angular radii are given as fraction of
the Fresnel scale both which are calculated at 40\,AU.}
\label{SI_fig2}
\end{figure}

\begin{figure} [htp]
\centerline{\includegraphics[ scale=1.7]{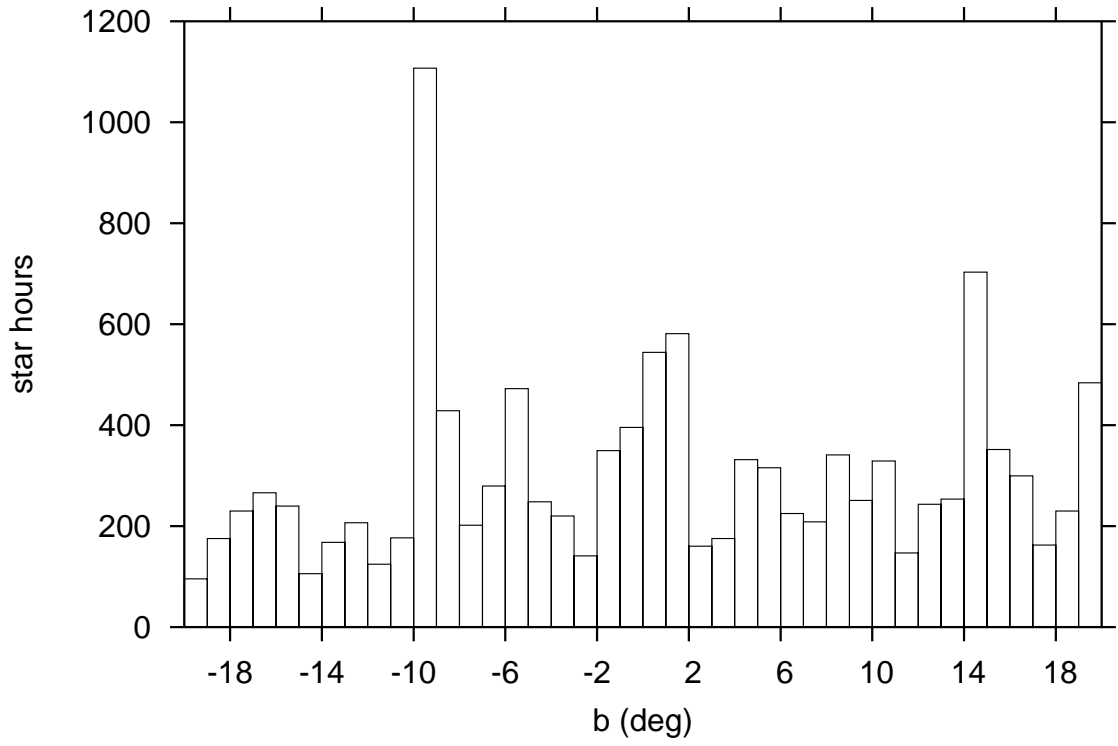}}
\caption{\footnotesize Distribution of star hours as a function of
  ecliptic latitude, $b$, for the 12,000 hours of low ecliptic latitude
  observations ($\vert b \vert < 20^{\circ}$) in the analyzed FGS data set.}
\label{SI_fig3}
\end{figure}

\begin{figure} [htp]
\centerline{\includegraphics[ scale=1.7]{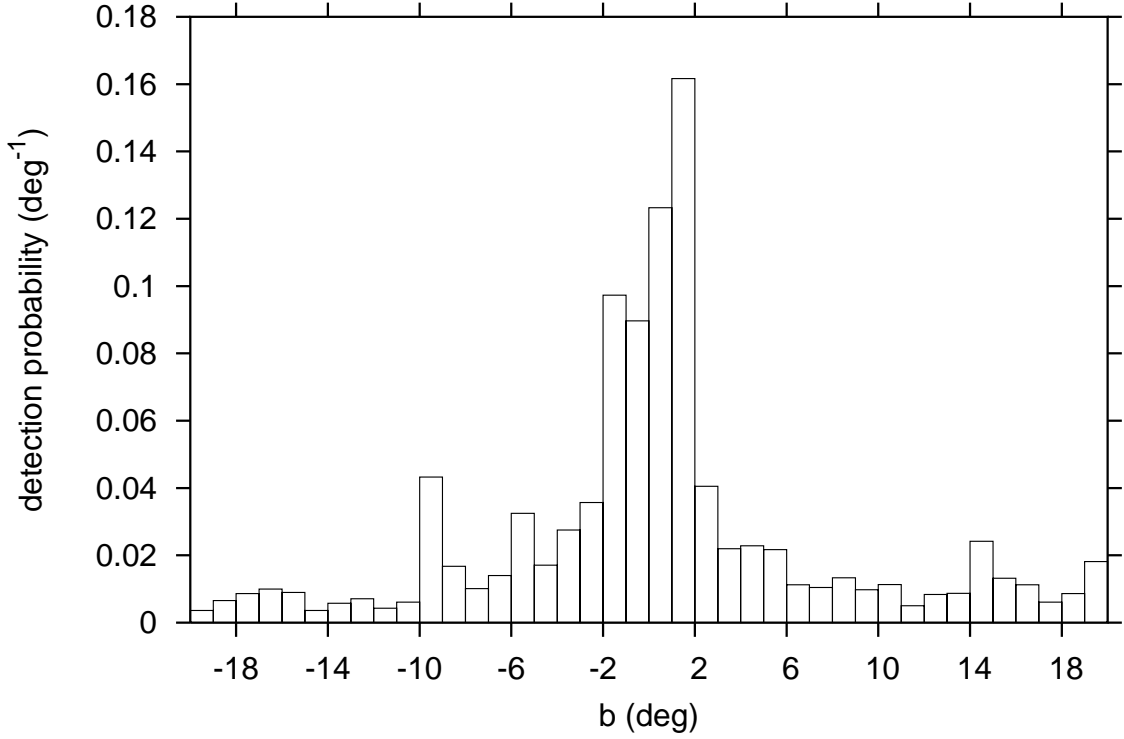}}
\caption{\footnotesize Detection probability as a function of ecliptic
  latitude, $b$, for the 12,000 hours of low ecliptic latitude observations
  ($\vert b \vert < 20^{\circ}$) of the analyzed FGS data set. The detection
  probability was calculated from the ecliptic latitude distribution of FGS
  guide stars shown in Supplementary Figure \ref{SI_fig3} and the KBO ecliptic
  latitude distribution from Elliot et al. (2005)\cite{E052}. Note, we assumed
  that the KBO ecliptic latitude distribution is symmetric about the ecliptic
  and ignored the small $\sim 1.6^{\circ}$ inclination of the Kuiper belt
  plane\cite{E052} relative to the ecliptic. For our survey, there is $\sim
  60$\% probability that KBO occultations will occur inside the
  low-inclination core region ($\vert b \vert < 4^{\circ}$) of the Kuiper
  belt. The probability for KBO occultations outside the core region is
  roughly uniform for $4^{\circ}<\vert b \vert < 20^{\circ}$ and about 40\% of
  all KBO occultations will occur outside the low-inclination core
  region. The detection of one object at $14^{\circ}$ is therefore consistent
  with the latitude distribution of our observations and that of KBOs.}
\label{SI_fig4}
\end{figure}

\begin{figure} [htp]
\centerline{\includegraphics[ scale=1.7]{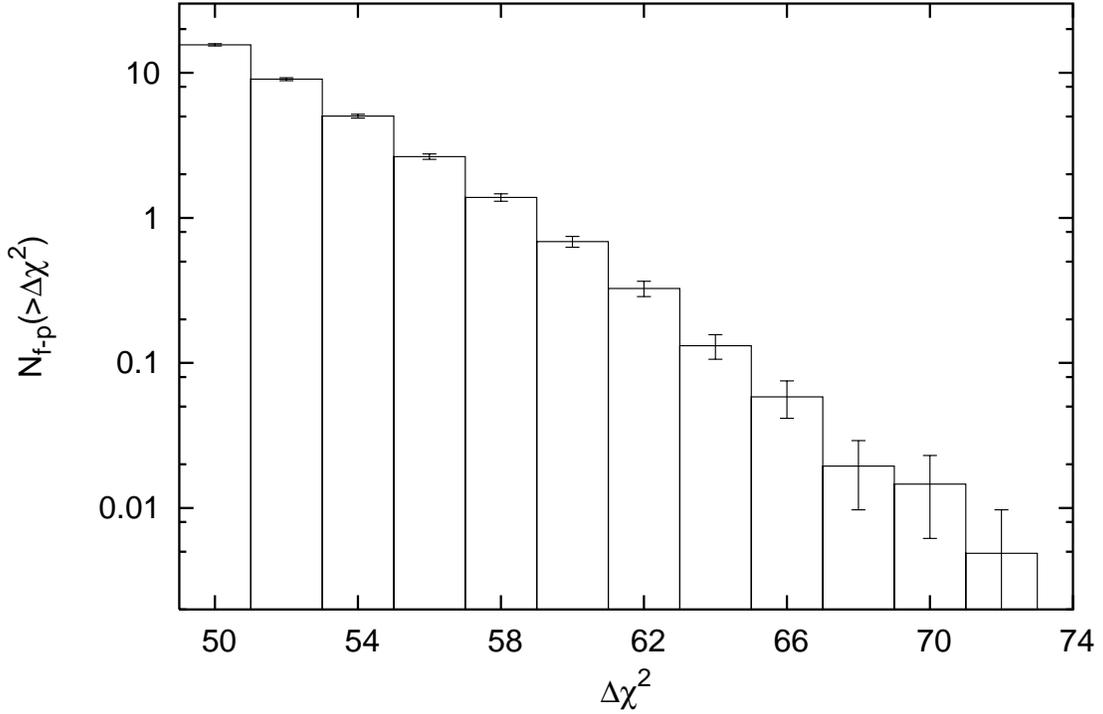}}
\caption{\footnotesize Cumulative number of false-positives, $N_{f-p}$, as a
  function of $\Delta \chi^2$. These false-positives were obtained from
  bootstrap simulations using data from $\sim 28$ minutes of \fgs observations
  that were acquired over one \hst orbit, in which we discovered the
  occultation candidate. The original time series was 32 minutes long and we
  removed the last 4 minutes that showed a significant increasing trend in the
  number of photon counts. We removed the occultation event itself (which
  occurred about 2.3 minutes before the start of the trend) and simulated $2.5
  \times 10^{6}$ star hours, which is 206 times larger than our low ecliptic
  latitude observations. This calculation required $\sim 1400$\,CPU days of
  computing power. The number of false-positives, $N_{f-p}$, was normalized to
  12,000 star hours, which corresponds to the length of the entire low
  ecliptic latitude observations. In the entire bootstrap analysis we obtained
  4 events with a $\Delta \chi^2 \ge 67.3$. This implies a probability of $8
  \times 10^{-7}$ that events like the occultation candidate with $\Delta
  \chi^2 = 67.3$ are caused by random statistical fluctuations within the
  original 32 minutes data set that contained the event and a probability of
  $\sim 4/206 \sim 2\%$ that events like the occultation candidate are caused
  by random statistical fluctuations over the entire low ecliptic latitude
  observations. The analysis of our high ecliptic latitude control sample,
  which is twice as large, did not yield any events that were comparable in
  significance to the occultation candidate.}
\label{SI_fig5}
\end{figure}

\end{document}